\documentclass[11pt, letterpaper]{article}

\usepackage{amsfonts, amsmath, amsthm, amssymb, graphicx, verbatim}
\usepackage{algorithmic, algorithm}
\usepackage[margin=1.2in]{geometry}
\usepackage{booktabs} 
\usepackage[font=small,labelfont=bf]{caption} 
\usepackage{amsfonts, amsmath, amsthm, amssymb, graphicx, verbatim} 
\usepackage{wrapfig} 
\usepackage{tabularx}
\usepackage[colorinlistoftodos,textsize=tiny]{todonotes} 
\usepackage{bbm}
\usepackage{thm-restate}
\usepackage{natbib}

\newlength{\bibitemsep}
\newlength{\bibparskip}
\let\oldthebibliography\thebibliography
\renewcommand\thebibliography[1]{%
  \oldthebibliography{#1}%
  \setlength{\parskip}{\bibitemsep}%
  \setlength{\itemsep}{\bibparskip}%
}

\usepackage{float}
\usepackage{enumitem}

\newcommand{\Comments}{0}
\newcommand{\mynote}[2]{\ifnum\Comments=1\textcolor{#1}{#2}\fi}
\newcommand{\mytodo}[2]{\ifnum\Comments=1%
	\todo[linecolor=#1!80!black,backgroundcolor=#1,bordercolor=#1!80!black]{#2}\fi}

\begin{document}

\title{Bridging Machine Learning and Mechanism Design towards Algorithmic Fairness}

\author{Jessie Finocchiaro\\
CU Boulder \\
\and
Roland Maio \\
Columbia University \\
\and
Faidra Monachou \\
Stanford University \\
\and
Gourab K Patro \\
IIT Kharagpur \\
\and
Manish Raghavan \\
Cornell University \\
\and
Ana-Andreea Stoica\\
Columbia University\\
\and
Stratis Tsirtsis \\
Max Planck Institute for Software Systems
}

\maketitle
\begin{abstract}
Decision-making systems increasingly orchestrate our world: how to intervene on the algorithmic components to build fair and equitable systems is therefore a question of utmost importance; one that is substantially complicated by the context-dependent nature of fairness and discrimination.
Modern decision-making systems that involve allocating resources or information to people (e.g., school choice, advertising) incorporate machine-learned predictions in their pipelines, raising concerns about potential strategic behavior or constrained allocation, concerns usually tackled in the context of mechanism design.
Although both machine learning and mechanism design have developed frameworks for addressing issues of fairness and equity, in some complex decision-making systems, neither framework is individually sufficient.
In this paper, we develop the position that building fair decision-making systems requires overcoming these limitations which, we argue, are inherent to each field.  
 Our ultimate objective is to build an encompassing framework that cohesively bridges the individual frameworks of mechanism design and machine learning.
We begin to lay the ground work towards 
this goal by comparing the perspective each 
discipline takes on fair decision-making, teasing out the lessons each field has taught and can teach the other, and highlighting application domains that require a strong collaboration between these disciplines.
\end{abstract}

\section{Introduction}\label{sec:intro}

Centralized decision-making systems are being increasingly automated through the use of algorithmic tools: user data is processed through algorithms that predict what products and ads a user will click on, student data is used to predict academic performance for admissions into schools and universities, potential employees are increasingly being filtered through algorithms that process their resume data, and so on.
Many of these applications have traditionally fallen under the umbrella of mechanism design, from auction design to fair allocation and school matching to labor markets and online platform design.
However, recent pushes towards data-driven decision-making have brought together the fields of mechanism design (MD) and machine learning (ML), creating complex pipelines that mediate access to resources and opportunities. 
Increasingly, learning algorithms are used in the context of mechanism design applications by adopting reinforcement learning techniques in auctions \citep{feng2018deep,dutting2019optimal,zheng2020ai, tang2017reinforcement} or general machine learning algorithms in combinatorial optimization \citep{bengio2020machine} and transportation systems \citep{jones2018future}. As such applications do not directly focus on fairness and discrimination, they are not the central focus of this paper.

The growing impact of these decision-making and resource allocation systems has prompted an inquiry by computer scientists and economists: 
are these systems fair and equitable, or do they reproduce or amplify discrimination patterns from our society?
In building fair and equitable systems, the question of fairness and discrimination is often a contested one.
Paraphrasing \citet{dworkin2002sovereign}, \textit{``People who praise or disparage [fairness] disagree about what they are praising or disparaging.''}
The causes of these philosophical debates include divergent value systems and the context-dependent nature of fairness and discrimination.
However, even when we do agree on the types of harms and discrimination we seek to prevent, mechanism design and machine learning often provide different sets of techniques and methodologies to investigate and mitigate these harms.
A key goal of this work is to identify the gaps between how machine learning and mechanism design reason about how to treat individuals fairly and detail concrete lessons each field can learn from the other.
Our hope is that these lessons will enable more comprehensive analyses of joint ML--MD systems.

Where do the gaps between machine learning and mechanism design come from?
Crucially, each field tends to make assumptions or abstractions that can limit the extent to which these interventions perform as desired in practice.
This limitation is not specific to machine learning and mechanism design; in general, any field must choose an appropriate scope in which to operate, i.e., a \textit{reducibility assumption}: it is assumed that the issue at hand is reducible to a standard domain problem, and that if the solution to this problem is fair and equitable, then so too will be the overall sociotechnical system \citep{selbst2019fairness}.
Under the reducibility assumption, fairness and discrimination can be addressed by an intervention that operates within the frame of the field in question, whether that be a constraint on a machine learning algorithm or a balance between the utilities of various agents in a mechanism.
Yet, in practice, complex algorithmic decision-making systems rarely satisfy any sort of reducibility assumption; not only do these systems require the combination of ideas from both disciplines, 
they also depend heavily on the social and cultural contexts in which they operate.

Our goal here is not to argue that it is \textit{sufficient} to consider machine learning and mechanism design in conjunction with one another; rather, we argue that it is \textit{necessary} to do so.
Working within each field in isolation will ultimately lead to gaps in our broader understanding of the decision-making systems within which they operate, making it impossible to fully assess the impact these systems have on society.
Of course, broadly construed sociotechnical systems cannot be fully understood just through these technical disciplines; our hope is that a more robust understanding of the strengths and weaknesses of machine learning and mechanism design will allow for a clearer view into how they can be integrated into a broader study of these sociotechnical systems.

As an illustrative example, consider the problem of online advertising.
Most modern online advertising systems perform a combination of prediction tasks (e.g., how likely is a user to click on this ad?) and allocation tasks (e.g., who should see which ad?).
Moreover, these advertising systems significantly impact individuals' lives, including their access to economic opportunity, information, and housing, and new products and technologies (see, e.g.,  \cite{facebookhud, facebookaddiscrimination}).
Thus, advertising platforms must consider the social impact of their design choices and actively ensure that users are treated fairly.

In isolation, techniques to ensure fair ad distribution from either machine learning or mechanism design fail to fully capture the complexity of the system.
On the mechanism design side, auctions typically take learned predictions as a given; as a result, they can overlook the fact that algorithmic predictions are trained on past behavior, which may include the biased bidding and targeting decisions of advertisers.
On the other hand, while evaluation tools from fair machine learning would help to ensure that the predictions of interest are ``good'' for everyone (by some definition), they may fail to capture the externalities of competition between ads that might lead to outcome disparities  \citep{ali2019discrimination}.
For example, a job ad may be shown at a higher rate for men than for women because it must compete against a different set of ads targeted at women than at men.
As each field has only a partial view of the overall system, it might be impossible to reason about the system's overall impact without taking a broader view that encompasses both the machine learning and mechanism design considerations.

This disconnect is not limited to the ad auction setting described above.
Due to their historically different applications and development, both machine learning and mechanism design tend to make different sets of assumptions that do not always hold in practice, especially in pipelines that combine tools from both fields.
On the one hand, machine learning traditionally treats people as data points without agency and defines objectives for learning algorithms based on loss functions that depend either on deviations from a ground truth or optimize a pre-defined metric on such data points. Thus, machine learning definitions of fairness tend to ignore complex preferences, long-term effects, and strategic behavior of individuals. On the other hand, as mechanism design often assumes known preferences, and more generally, that information comes from a fixed and known distribution without further critique, and measures utility as a proxy for equality, it tends to miss systematic patterns of discrimination and human perceptions (see also Section \ref{sec:humanperceptions}). While recent works have started to address these gaps between machine learning and mechanism design approaches to fairness by embedding welfare notions in measures of accuracy and fairness and using learning algorithms to elicit preferences, many open questions remain on what each field can learn from the other to improve the design of automated decision-making systems.

In this paper, we formalize these ideas into a set of lessons that each field can learn from the other in order to bridge gaps between different theories of fairness and discrimination. 
In doing so, we aim to provide concrete avenues to address some of the limitations of machine learning and mechanism design, under the acknowledgement that bridging these fields is only an initial step towards a comprehensive analysis of sociotechnical systems.

We make the following contributions: 

\begin{itemize}[noitemsep,topsep=0pt]
    \item We review definitions of fairness and discrimination in machine learning and mechanism design, highlighting historical differences in the way fairness has been defined and implemented in each (Section~\ref{sec:differences}).
    \item We define several lessons that can be learned from mechanism design and machine learning in order to create an encompassing framework for decision-making.
    Specifically, we highlight the gap between fairness and welfare, the potential of long-term assessment of decision making systems, group versus individual assessment of fairness and the effect of human perception of fairness, among others (Section~\ref{sec:pastfuturelessons}).
    \item Finally, we highlight different application domains and survey relevant works in which both mechanism design and machine learning tools have been deployed, such as  advertising, education, labor markets and the gig economy, criminal justice, health insurance markets, creditworthiness, and social networks. We discuss advances and limitations of current techniques and implementations in each of these domains, relating to the lessons from the previous section (Section~\ref{sec:applications}).
\end{itemize}

\section{Differences between Mechanism Design and Machine Learning}\label{sec:differences}

Machine learning has been increasingly used to supplement human decisions, drawing attention to biases rooted in learning from historically prejudiced data \citep{angwin2016machine,buolamwini2018gender,barocas2016big}.
Fair machine learning often defines fairness conditions (e.g. parity for legally protected groups) without considering core mechanism design concerns such as welfare and strategic behavior.
Yet, mechanism design often fails to conceptualize the impact of decisions for different social groups.

While both fields incorporate quantitative notions of fairness into optimization, they differ in the roles those notions play: in machine learning, fairness is typically a constraint to be satisfied, hence the learning algorithms are not optimizing for the \emph{most} fair solution; in contrast, mechanism design typically defines and directly optimizes a fair utility-based objective (e.g., social welfare).

This is only one of many high-level differences between the two fields.
 \citet{abebe2018mechanism} and  \citet{kasy2020fairness} indirectly observe that understanding those differences and bridging  different notions of fairness is essential in improving access to opportunity for different communities, as well as extending the purpose of each field to encompass the causal effect of algorithmic design on inequality and distribution of power  \citep{kasy2020fairness}.

\subsection{Fairness in machine learning}\label{subsec:fair-ml}
Multiple definitions of fairness have been proposed; 
interestingly, their common characteristic seems to be that they agree to disagree.
 \citet{mehrabi2019survey} collect 
the most common fairness definitions; 
most of them
fall into two main categories, \emph{individual} and \emph{group} fairness.
%
Group fairness notions assess the large-scale effect of an algorithmic system on different demographic groups (often defined by legally protected classes).
Individual fairness, however, compares outcomes between each individual in a population, requiring people who are similar to each other to receive a similar outcome, and is therefore typically a stronger constraint.

\vspace{0.37em}
\noindent\textbf{Individual fairness.}
Inspired by  \citet{rawls2009theory}' fair equality of opportunity  in political philosophy, 
 \citet{dwork2012fairness} formalize the notion of individual fairness as a constraint in a classification setting where one wants to \textit{``treat similar individuals similarly''} based on a pairwise similarity metric of their features, (partially) designed by domain experts.  
However, defining similarity metrics is not easy, especially between individuals belonging to sub-populations with different characteristics.
Subsequent work, though limited, mainly aims to overcome this obstacle by either learning feature representations that conceal the individuals' membership to a protected
group \citep{zemel2013learning, lahoti2019ifair} or by selecting individuals based on how they compare in terms of qualification with other members of their own sub-group \citep{kearns2017meritocratic}. However, individual fairness is not equivalent to meritocracy, since qualified covariates might be more difficult to obtain for disadvantaged people, meaning one person may have 
worked harder to be recognized as ``similar'' by the algorithm \citep{hu2019disparate}.

Overall, individual fairness is reminiscent, yet different, from the individual perspective that utility measures in mechanism design often take (e.g., the notion of envy-freeness from mechanism design) and can be used to compare metrics that assess the individual experience in an algorithmic setting. While individual fairness does not take into account one's preferences (often assumed in mechanism design), recent works \citep{kim2019preference} re-design this definition by taking into account individual preferences.

\vspace{0.37em}\noindent\textbf{Group fairness.}
Numerous definitions have been proposed for group fairness \citep{verma2018fairness,mehrabi2019survey}, suggesting to impose either simple statistical parity conditions between groups \citep{ corbett2017algorithmic}
or more complex classification constraints; some 
aim to equalize each group's opportunity to positive outcomes \citep{hardt2016equality},  balance the misclassification rates among groups \citep{berk2018fairness, zafar2017fairness} or provide similar classifications under counterfactual group memberships \citep{kusner2017counterfactual}.
 \citet{kleinberg2017inherent} and  \citet{chouldechova2017fair} show that tensions arise when trying to simultaneously achieve 
multiple notions. 
However, as  \citet{madaio2020co} emphasize, if one is to strive for quantitative fairness, the notion one optimizes for should be context-dependent and developed in partnership with stakeholders.

\vspace{0.37em}
 Despite the variety of individual and group fairness definitions, it  becomes apparent that  they lack \textit{expressiveness}. Most of these definitions focus solely on the inputs and outputs of the algorithm without taking into account how those outputs ultimately impact real-world outcomes.
For example, the most common assumption is that a ``positive classification'' output is an equally valuable outcome for everyone. As we discuss in Sections~\ref{subsec:fair-md} and~\ref{sec.tension_fairness_welfare}, mechanism design can offer the tools and definitions to overcome such limitations and successfully incorporate important aspects such as  individual- and group-level utilities,  resource constraints, as well as strategic incentives, to the design of decision-making models.

\subsection{Fairness in mechanism design}
\label{subsec:fair-md}

The mechanism design literature shifts the focus away from \emph{fairness} towards \textit{welfare} and \emph{discrimination.}
We review (i) the classic theories of taste-based and statistical discrimination, (ii) utilitarianism and the idealized objective of maximum social welfare, and (iii) fairness in social choice theory.

\vspace{0.37em}\noindent\textbf{Economic theories of discrimination.}
There are two prevalent economic theories of discrimination: \textit{taste-based} and \textit{belief-based}.
The key difference between them is the effect of information;
taste-based discrimination arises due to pure preferences  \citep{becker1957economics}, and persists even with perfect information about individuals. 
This theory has often been criticized as being simplistic
since it is based on the discriminatory principle that decision-making agents derive higher utility from certain social groups  \citep{guryan2013taste}; however, empirical evidence is rather inconclusive and application-dependent  \citep{charles2008prejudice, altonji2001employer, knowles2001racial, cui2017discrimination}.

The latter theory of belief-based discrimination can be particularly informative for the design of fair machine learning systems as the true attribute of an agent is often not observed directly, but only through a proxy. From this theory, \textit{statistical discrimination} \citep{arrow1973theory,phelps1972statistical} generally assumes that differences are exogenous but exist. Other papers attribute discrimination to \textit{coordination failure}: agents are born unqualified but can undertake some costly skill investment, which may lead to asymmetric equilibria  \citep{coate1993will}. 
Finally, another  belief-based discrimination theory is \textit{mis-specification}  \citep{bohren2019dynamics};
unaware of their own bias \citep{pronin2002bias},  decision-makers may hold misspecified models of group differences which, in the absence of perfect information, lead to false judgment of an individual's abilities.

Such economic models  offer useful insights on how to design a system aware of inequality due to (i) equilibrium asymmetries, (ii) information limitations, and (iii) human behavioral biases. 
For example, different social groups may differ in their skill level due to systematic inequalities of opportunity when certain equilibria arise, but not due to inherited differences in their true ability. 
This may be in sharp contrast to human decision-makers (or even algorithms) who, due to imperfect information or other biases, may incorrectly infer that perceived differences among individuals can be perfectly explained by observed characteristics. 

\vspace{0.37em}\noindent\textbf{Utilitarianism and normative economics.}
Beyond discrimination theories, utilitarianism and normative economics have been extensively used in mechanism design to motivate using utility functions as a synonym for social welfare. Although these two terms are used interchangeably and welfare economics is often viewed as applied utilitarianism, their origin differs.
As  \citet{posner1983economics} writes, \textit{utilitarianism} is a philosophical system which holds that \textit{``the moral worth of an action, practice institution or law is to be judged by its effect on promoting happiness of society.'' } 
On the other hand, \textit{normative} or \textit{welfare economics} holds that \textit{``an action is to be judged by its effects in promoting the social welfare.''}  \citep{posner1983economics} 
In contrast to machine learning and its multiple definitions of fairness,  weighted social welfare is the most  accepted measure of broader ``social good'' in mechanism design but not necessarily of fairness or equity. 
Typically, utilitarian approaches capture equity by assigning appropriately defined weights to the utility of each agent. 
Nevertheless, a major limitation remains as welfare economics models rarely explain how to come up with these weights and how to interpret the relative difference between two agents' weights.

\vspace{0.37em}\noindent\textbf{Fairness in social choice theory.} 
Social choice theory deals with collective decision making processes, and fairness is of great significance in such processes---particularly in resource allocation problems and voting.
In fair allocation, the goal is to divide a resource or set of goods among $n$ agents that is somehow ``fair.''
The literature tends to focus on three primary notions of fairness:
{\it proportional division}  \citep{steihaus1948problem} (every agent receives at least $\frac{1}{n}$ of her perceived value of resources);
{\it equitability}  \citep{foley1967resource} (every agent equally values their allocations);
and {\it envy-freeness}  \citep{varian1973equity} (every agent values their allocation at least as much as another's).
While these notions capture fairness of allocations at an individual level, they treat all  individuals equally 
in contrast to individual fairness which relies on some similarity metric to ensure similar outcomes only for similar
individuals.
Moreover, in many real-world problems in healthcare, finance, education, relaxed notions of fairness  are used due to the hardness of the absolute notions.
 \citet{conitzer2019group} points out that one deficiency of relaxed notions of fair allocation is that they fail to capture group-level disparities and often leave room for group unfairness (see Section \ref{subsec:redefining_fairness}).
Finally, another difference is that, unlike machine learning settings where all individuals prefer positive or higher outcomes, social choice theory can naturally capture different preferences of agents over the possible outcomes.\footnote{Voting theory deals with aggregating  individual preferences.  
We exclude discussions on voting while we acknowledge the existence of substantial works on fair voting.}

\section{Past and Future Lessons}\label{sec:pastfuturelessons}

We enumerate several lessons that mechanism design (MD) and machine learning (ML) are able to learn from each other. 
We denote by $A \to B$ a lesson that has been or can be taught by field $A$ to  $B$.

\subsection{MD $\to$ ML: Tension between fairness and welfare}
\label{sec.tension_fairness_welfare}

 \citet{kaplow2003fairness} are among the first to argue, from a legal and economic point of view, that \textit{``the pursuit of notions of fairness results in a needless and, at root, perverse reduction in individuals’ well-being,''} and that welfare should be instead the primary metric for the effectiveness of a social policy. 
Optimizing for fairness instead of welfare can actually cause harm in social decision-making processes (e.g., by leading to a violation of the Pareto principle).
This is later supported for quantitative fairness metrics by \citet{hu2020fair,hossain2020designing}, who show that adding group parity constraints can decrease welfare for \emph{every} group. 

Recent works also propose fairness-to-welfare pathways 
that transform utility-based metrics into comparing probability of outcome \citep{zafar2017parity,balcan2019envy,hossain2020designing},   showing that fairness definitions do not automatically imply equitable outcomes from a mechanism design perspective, but on the contrary. 
 \citet{kasy2020fairness} formalize some of these tensions, arguing that machine learning definitions fail to acknowledge inequality within protected groups as well as perpetuate it through notions of merit. 
This is further complicated by the fact that, while notions of fairness in machine learning often treat outcomes as binary with a single desirable outcome, the real world is  far more complex; different individuals may have different preferences over a wide range of outcomes. While individual fairness is often incorporating stronger constraints to ensure that individuals receive a good outcome given their features, their preferences are not directly taken into account. Recent works are addressing this gap by re-designing notions of fairness with preferences in mind \citep{kim2019preference}.

Using the lens of welfare economics as well as economic theories of discrimination to assess the equitability of machine learning systems can be useful for designing just systems, but it is no panacea.
An important question that arises is whether the prevalent utilitarian view of mechanism design is already problematic. A common criticism of utilitarianism is that it is not clear whose utilities we should maximize and how much weight each individual should receive in the optimization objective. 
For example, should an algorithm ensure the average utilities of both protected and unprotected groups be the same, or should each group contribute to the total welfare proportionally to its size in society? 
If we search beyond economics and computer science, we soon realize that practical difficulties and tensions in philosophy, political science, history, sociology and other disciplines are similar to some of the tensions we currently see in machine learning. 
For example, borrowing from political philosophy,  \citet{binns2017fairness} introduces new notions of fairness  that challenge both the common concept of social welfare maximization and fair machine learning definitions, by asking questions such as:  
\textit{should we minimise the harms to the least advantaged?}
In the end, while there may be no universal notion of welfare that adequately captures society's beliefs about whose welfare to prioritize, mechanism design provides the tools to begin to interrogate these welfare trade-offs in a way that machine learning has yet to fully reckon with.

\subsection{MD $\to$ ML: Long-term effects of fairness}\label{sec:longterm}

Because mechanism design considers outcomes for an entire population of agents, the machine learning community has started to adopt mechanism design techniques (ranging from 
equilibria analysis in games to dynamic models of learning agents) in order to study the effects of machine learning algorithms on different subpopulations.
For example, 
the decisions made by an algorithm and the (strategic) participants can change the population data over time, requiring learning to be dynamic rather than one-shot. 

Economics has long studied such dynamic effects, but without a machine learning perspective. 
However, several useful lessons can be extracted from recent works \citep{zhang2020fairness}. 
First and foremost, dynamic effects over time are crucial, and, if neglected, they can worsen rather than improve  inequality and discrimination in large-scale decision-making systems. 
Indeed, even simple two-stage models show that it is impossible to achieve full equality and have the potential of causing harm due to fairness constraints  \citep{liu2018delayed,kannan2019downstream}; interestingly, such models and subsequent works \citep{liu2020disparate} are strongly influenced by the classic economic models such as  \citet{coate1993will} and  \citet{phelps1972statistical}. 

Second, the type and complexity of interventions needed to achieve long-term fairness may vary significantly. 
For example,  \citet{hu2018short} build upon the labor market model in  \citet{levin2009dynamics} and showcase the positive effect of simple short-term restrictions (via a group demographic parity constraint) on improving long-term fairness. However, other systems may require a more complex approach;  \citet{wen2019fairness} study fairness in infinite-time dynamics by using a Markov Decision Process to learn a policy for decision-making that achieves demographic parity or equalized odds in the infinite time dynamics. From a technical perspective, increasing leaning on popular mechanism design tools such as large market models, mean-field equilibria analysis, and dynamic programming techniques seems to be a promising direction for the design of effective and fair policies in machine learning-driven systems.

Finally, most machine learning models focus solely on algorithmic bias and are oblivious to the existence of the social bias that is coming from human agents making complex, dynamic decisions as a response to the system's algorithmic decisions. 
The interplay between social and algorithmic bias over time  may in fact prove itself useful in explaining dynamic patterns of discrimination in sociotechnical systems. 
 \citet{bohren2019dynamics} introduce the discrimination theory of mis-specification and 
 show, both theoretically and empirically, that contradicting patterns of discrimination against women's evaluations in online platforms can be well explained by users' mis-specified bias in sequential ratings.
 \citet{monachou2019discrimination} build upon this theory and tools for learning from reviews 
to  study the long-term effects of social bias on worker welfare inequality in online labor markets, while   \citet{heidari2019long} also use observational learning to study the temporal relation between social segregation and unfairness.

\subsection{MD $\to$ ML: Strategic agents}

The economist's basic analytic tool is the assumption that people are \textit{rational maximizers} of their utility, and most principles of mechanism design are deductions from this basic assumption.
Therefore, as machine learning algorithms are increasingly used in prescriptive settings, like hiring or loan approval, it becomes necessary to consider the incentives of the agents who are affected from those algorithmic decisions.
As transparency laws regarding algorithmic decision-making are gradually being introduced \citep{voigt2017eu}, individuals are now more than ever capable to use insights about the deployed classifiers and accordingly alter their features in order to ``game'' the system and receive a beneficial outcome.

This observation has initiated a line of work on \textit{strategic classification} \citep{dalvi2004adversarial,bruckner2011stackelberg,bruckner2012static, hardt2016strategic, dong2017strategic, chen2020learning, hu2019disparate} which focuses on incentive-aware machine learning algorithms that try to reduce misclassification caused by transparency-induced strategic behavior.
The ability to manipulate their features naturally raises several fairness questions.
For example,  \citet{hu2019disparate} contextualize strategic investment in test preparation to falsely boost scores that are used as a proxy to quantify college readiness, as well as the disparate equilibria that could potentially emerge in the presence of social groups with disproportionate manipulation capabilities.
Additionally,  \citet{milli2019social} utilize credit scoring and lending data to show that there is a trade-off between the utility of a decision-maker who tries to protect themselves from the agents who modify their features strategically and the social burden different groups of agents incur as a consequence.

On a more positive note, recent work has argued that this strategic modification of features does not always correspond to an agent's attempt to ``game'' the system but could also represent a truthful investment of effort towards improvement, depending on the features being used and the extent to which they can be maliciously manipulated.
This idea has become apparent both in the mechanism design literature \citep{kleinberg2019classifiers, alon2020multiagent} on evaluation mechanisms and the machine learning literature \citep{tsirtsis2020optimal, miller2019strategic, haghtalab2020maximizing} on the design of transparent decision policies that aim to incentivize the individuals' improvement.
Relaxing our initial assumption about strict individual rationality, we can easily see that transparent decision policies based on features prone to manipulation may prove themselves substantially unfair, by equally rewarding seemingly similar individuals with dissimilar effort profiles (in direct opposition to definitions of individual fairness), as those dissimilarities may have ethical, behavioral or cultural origins.
For ease of exposition, consider a simple example of admitting graduate students solely based on their undergraduate GPA.
Even if two students share the same observable features (GPA), that could reflect different mixtures of manipulating the undergraduate evaluation rules or achieving truthful academic excellence, a behavior often depending on their cultural background \citep{magnus2002tolerance,payan2010effect}.  
In this context, the uncertain relation between features and individual qualifications gives rise to a need for \textit{strategyproofness} in order to make prediction-based decision-making systems transparent and  fair.

Apart from simple classification settings, the interplay between machine learning and mechanism design also needs to be considered in more complex systems where the stakeholders have more diverse incentives and predictive models of different forms also appear.
For example, in health insurance markets machine learning is used to predict the expected costs of individuals and proportionally compensate insurers, with strategic upcoding by the latter favorably skewing subsequent predictions \citep{cunningham2012risk} and disincentivizing all insurers from offering attractive insurance plans to people with specific medical conditions \citep{zink2020fair}.
Moreover, the retrieval and recommender systems, well-known downstream applications of machine learning, are also vulnerable to strategic behavior leading to disparate effects even in the absence of model transparency;
specifically, strategic manipulation in recommendations \citep{chakraborty2019equality,song2020poisonrec} and search engines \citep{baruchson2007manipulating,epstein2015search} often results in skewed information delivery leading to disproportional opportunity or exposure for the users.
Such disparate effects of machine learning highlight the need for further research towards the direction of developing models aware of the strategic environment in which they operate as well as the effects of their predictions on different people and groups.

\subsection{ML $\to$ MD: Defining and diagnosing unfairness under uncertainty}
\label{subsec:redefining_fairness}

Definitions of fairness from the mechanism design literature tend to be centered around preferences and utilities. 
As discussed earlier, the fair machine learning literature has yet to fully adopt this perspective, typically operating at the level of model outputs as opposed to the values for individuals produced by those outputs. 
However, a key assumption necessary for mechanism design’s preference-based notions of fairness is that individuals’ preferences are known or can be in some way communicated to a central decision-maker. 
In many mechanism design applications, like traditional auctions or school choice, this assumption can be reasonable. 
In more complex systems like online advertising, preferences are often unknown a priori and must be estimated in practice. 
Thus, questions of fairness necessarily involve reasoning about uncertainty and who bears the burden of errors. In this way, ideas about fairness from machine learning can be useful. 
Because machine learning treats uncertainty as a first class concept, many conceptions of fairness from the machine learning literature explicitly consider errors and their impact on different sub-populations \citep{hardt2016equality,chouldechova2017fair, zafar2017fairness}.

Uncertainty can also manifest itself with respect to outcomes, not just to preferences.
Many application domains utilize probabilistic models---for example, labor market models from mechanism design often consider two-stage processes in which noisy signals provide information about whether a worker is qualified or not \citep{coate1993will,hu2018short}. 
Importantly, while these models do incorporate uncertainty, the designer knows the true relationship between observed signals and true outcomes, even though this relationship is probabilistic. 
This style of analysis is less suited to deal with cases where the relationship between signals and ground truth is unknown and can only be learned about through data. The lack of ground-truth information greatly complicates any analysis of the impact of a mechanism, but it is precisely this lack of information that machine learning techniques are designed to handle. 
Many of the challenges that arise during learning, including data scarcity for certain groups \citep{buolamwini2018gender}, feedback loops \citep{ensign2018runaway}, preference elicitation \citep{zinkevich2003polynomial,blum2004preference,goldberg2020learning,frongillo2018axiomatic}, and explore-exploit trade-offs \citep{bird2016exploring,immorlica2019diversity,raghavan2018externalities}, implicate serious fairness concerns.
By integrating lessons from machine learning on how to define and measure disparities that learning produces, mechanism design can gain a deeper understanding of real-world systems.

Using fairness definitions as a diagnostic tool for potential harms and societal issues is a powerful application of computing, as \citet{abebe2020roles} argue. 
As such, the various group fairness definitions from machine learning focus on illustrating output differences between different legally protected groups, using error measurements to quantify such differences (e.g., false positive/negative rates). 
A single definition is thus not feasible, nor desirable, but the process of defining fairness has been expanding, both conceptually and practically: from early computer science works that defines fairness through observations \citep{dwork2012fairness,hardt2016equality} or representations \citep{zemel2013learning,feldman2015certifying} to understanding causal relationships between features \citep{kusner2017counterfactual,kilbertus2017avoiding}. 
While satisfying multiple definitions may not always be possible \citep{kleinberg2017inherent}, the different definitions of fairness in machine learning offer an opportunity to become more intersectional in defining sensitive groups and in assessing power differentials. 
More than that, they shift the purpose of defining fairness from a normative one to a diagnostic one, a purpose that mechanism design can learn from when assessing the utility of a system. 

Together with a plethora of works from economics that assess differences in welfare at a group level \citep{coate1993will,hu2018short}, recent works in mechanism design \citep{conitzer2019group} propose adapting individual notions of envy-freeness into group-level definitions through stability, e.g., no group of people should prefer the outcome of another group. 

The need to assess the outcome differences between groups becomes more pressing as machine learning tools are increasingly being used in traditional mechanism design applications, as previously discussed.
Recent works increasingly adapt group fairness methods inspired from machine learning to design fair voting procedures \citep{celis2017multiwinner} and advertising \citep{kim2019preference}, bridging the gap between the individual perspective of mechanism design methods and group-level definitions of fairness from machine learning. Beyond transferring lessons from machine learning to mechanism design, 
we argue that future design must encompass perspectives other than the purely computational one, from sociological understandings of harm and power to economic discrimination and theories of justice.

\subsection{ML $\to$ MD: Human perceptions and societal expectations of fairness} \label{sec:humanperceptions}

Early studies on fairness in both mechanism design and machine learning propose various mathematical formulations of fairness, and normatively prescribe how fair decisions should be made. 
However, given the impossibility to simultaneously satisfy multiple fairness notions  \citep{kleinberg2017inherent,chouldechova2017fair}, decision-making systems need to be restricted to only selected principles of fairness, a process that becomes challenging in certain applications, such as criminal justice, finance and lending, self-driving cars, and others.
Given such applications and their potential for harm, it is  essential for the chosen design and principles to be socially acceptable.
Thus, there is a need to understand how people assess fairness and how to infer societal expectations about fairness principles in order to account for all voices in a democratic design of decision-making systems.

A line of work \citep{woodruff2018qualitative, lee2018understanding, grgic2018human,  green2019disparate, srivastava2019mathematical, saha2020measuring} in machine learning research has taken steps towards this democratization goal through participatory sociotechnical approaches to fairness  \citep{baxter2011socio,van2012agent} by studying human perceptions and societal expectations of fairness.
Three major questions emerge from this line of work, which, we argue, are central in developing participatory mechanism design tools that incorporate preferences.
We discuss them next.

First, {\it whose perceptions or assessment of fairness should be considered?} 
While  \citet{awad2018moral} and  \citet{noothigattu2018voting} used crowdsourced preferences from lay humans in the famous moral machine experiment,  \citet{jaques2019moral} and  \citet{yaghini2019human} have argued that preferences should be taken only from relevant individuals (e.g., primary stakeholders, ethicists, domain experts), citing context-dependent aspect of fairness and the possible vulnerability of lay humans to societal biases.

Second, {\it what options and information should be made available to the participants?}
Some studies  \citep{harrison2020empirical,saxena2019fairness,awad2018moral,noothigattu2018voting} directly asked participants to choose the model with the best fairness notion or the best outcomes, whereas others  \citep{srivastava2019mathematical,grgic2018human,yaghini2019human} asked indirect questions to infer the acceptable fairness principles (e.g., whether they approve of certain differences in decision outcomes for pairs of individuals from different groups, or the overall outcome distribution). 
In a different approach,  \citet{grgic2018beyond} and  \citet{van2019crowdsourcing} study the validity of using certain input features in the decision-making process in order to achieve procedural fairness. 

Finally, {\it how should the individual preferences be aggregated?}
Even though most of the literature has followed some variant of majority rule for this,  \citet{noothigattu2018voting} and  \citet{kahng2019statistical} have argued for tools like score-based bloc voting or Borda count from voting theory  \citep{elkind2017properties} for better representation of participants' choices. These studies have also shown the need of model explainability  \citep{binns2018s,rader2018explanations,dodge2019explaining}, transparency  \citep{rader2018explanations,wang2020factors}, and context-specific feature selection  \citep{grgic2018beyond,van2019crowdsourcing} in improving societal fairness perceptions, which mechanism design has traditionally considered as out of scope or assumed to be known, leading to a recent surge in explainability and transparency studies in machine learning.
Future work in mechanism design can learn from such studies in challenging current assumptions about preferences, perceptions, and values. 

\section{Application Domains}\label{sec:applications}

In this section, we discuss several application domains of machine learning and mechanism design to illustrate the lessons of Section~\ref{sec:pastfuturelessons}, underscore the complex interplay between these domains, point out gaps, as well as potential ways of bridging these gaps.

We note that many of the applications are open to critique. 
One might object to the idea of deciding which students are qualified or unqualified to receive an education in college admissions. More fundamentally, one might argue that the overall social system (e.g., criminal justice) in which an application (e.g., recidivism prediction) is embedded is unjust, and further that this cannot be remedied by any technical fairness intervention.
We discuss applications merely as an illustration of the lessons we have articulated, and reiterate our position that it is necessary, though not sufficient, to bridge machine learning and mechanism design for algorithmic fairness.

\subsection{Online advertising}

Auction design (a subfield of mechanism design) deals with the optimal design of allocation and payment rules when a number of agents bid for a resource.
As online ad auctions run in a high-frequency online setup that demands automated and precise bidding from the agents, many ad platforms have deployed machine learning models to estimate  the relevance of an ad to a customer while using some high-level preferences about advertisers' budget, bidding strategies, and target audiences.
Using the automated bids derived from these relevance predictions, ad allocation mechanisms 
 \citep{ostrovsky2011reserve} are run to place specific ads every time a user visits a webpage,  
thus making the system a complex mix of interdependent components from both machine learning and mechanism design.

Recent studies show that the resulting ad delivery may be problematically skewed; users who differ on sensitive attributes such as gender  \citep{lambrecht2019algorithmic}, age  \citep{angwin2017facebook}, race  \citep{angwin2016facebook}, may receive very different types of ads. For example, search queries with Black-sounding names are highly likely to be shown ads suggestive of arrest records  \citep{sweeney2013discrimination}. In another study, women were shown relatively fewer advertisements for high-paying jobs than men with similar profiles  \citep{datta2015automated}. When ads are about housing, credit or employment, such disparities can harm equality of opportunity.

One cause of problematically skewed ad delivery is explicit targeting of users based on sensitive attributes  \citep{faizullabhoy2018facebook, angwin2016facebook}, which can be tackled by 
disallowing ad targeting based on sensitive attributes especially for housing, credit, and employment ads. 
Although major ad platforms like Google and Facebook had disallowed targeting of opportunities ad based on sensitive attributes, the advertisers could still exploit other personally identifiable information such as area code  \citep{speicher2018potential}, or using a biased selection of the source audience in the Lookalike audience tool by Facebook.
Following a lawsuit  \citep{spinks2019contemporary}, Facebook removed targeting options for housing, credit, and employment ads \citep{facebookaddiscrimination}.

Other studies  \citep{sapiezynski2019algorithms,ali2019discrimination} again reveal that ad delivery mechanisms could still result in skewed audience distribution based on sensitive attributes even in the absence of any inappropriate targeting.
These are often the results of competitive spillovers; relative competition between general opportunity ads and category-specific ads for items like women's fashion can result in opportunity ads being shown to more male audiences. 
This issue has been tackled from both advertisers’ side and auctioneer’s side. 
Solutions on the advertisers' side include running multiple ad campaigns for different sensitive groups (with parity-constrained budgets)  \citep{gelauff2020advertising}, or using different bidding strategies for different demographics groups \citep{nasr2020bidding}.
However, such type of targeting has been disallowed by the platforms because of earlier exploitation by discriminatory advertisers. 
Moreover, rational advertisers may not want to adopt solutions that decrease utility.
On the auctioneer’s side, the allocation mechanism can be redesigned to ensure fair audience distribution  \citep{dwork2018fairness, ilvento2020multi, chawla2020fairness, celis2019toward}. 
Along with the welfare optimization goal, 
group fairness constraints can be used to ensure fair audience distribution  \citep{celis2019toward}, and individual fairness   \citep{chawla2020fairness} or envy-freeness constraints  \citep{ilvento2020multi} can be adopted to ensure similar individual satisfaction of the users. 

Most of these papers have focused on the mechanism design of online ad delivery. 
Yet all components---advertisers’ strategies, platform’s relevance prediction, ad allocation mechanism---may be responsible for unfair ad delivery.
While the mechanism design components take the relevance predictions from machine learning models as inputs, they often overlook the possibility of biases in these predictions.
Thus, to build a fair online ad ecosystem, there is a need to study the role of relevance prediction models and their role in the mechanism design pipeline.
In this regard, a line of work in machine learning that studies preference elicitation in auction settings  \citep{parkes2005auction,zinkevich2003polynomial,lahaie2004applying} 
can be explored and extended to online advertisements.

\subsection{Admissions in education} 

Schools and universities  increasingly use machine learning
to  inform admissions decisions \citep{clemsonadmissions}. 
Mechanism design has traditionally studied problems such as school choice, college admissions and affirmative action (e.g.,  \citep{abdulkadirouglu2003school,chade2014student,abdulkadirouglu2005college, chan2003does, fu2006theory, kamada2019fair, foster1992economic, immorlica2019access}). 
In general, most of these papers adopt similar assumptions and approaches.
At their baseline, they model the problem as a two-sided ``market'' of strategic agents: schools or colleges on the one side and students on the other side. In school choice, the assignment decisions are usually centralized (e.g., all public schools in Boston 
may commit to a common matching process), while in college admissions, 
each university decides independently which applicants to admit. 

In both cases, explicit fairness considerations are rarely taken into account. The only exception is, of course, \textit{affirmative action}, which is imposed as an additional external constraint on the market. Most economics papers have mostly considered two  categories of  policies with respect to protected attributes: \textit{group-unaware}

and \textit{group-aware} policies  \citep{fang2011theories}. Both policy schemes usually translate to demographic parity constraints and similar quota rules.

Interestingly, explicit notions of fairness and equity are less commonly considered.
This may be due to various reasons. For example, in a decentralized system such as college admissions, it is unclear whether and---most importantly---how to optimize social welfare. But even in more centralized applications,  such as school choice, 
several dilemmas arise. Given that both market sides  have heterogeneous preferences and strategic incentives,  should the central planner prioritize the students' or schools' welfare? How is social welfare even practically defined in this case?
Indeed, several papers  \citep{roth2008deferred,pathak2017really,robertson2020if,hitzig2018bridging,li2017ethics} have offered a broader critique of the approaches used by market designers, pointing to the gap between translating  theoretical assumptions to practical solutions.

Machine learning algorithms are increasingly being used in this area as well, for the purpose of parsing data at a large scale more efficiently and embedding missing notions of fairness. 
The machine learning literature  \citep{haghtalab2020maximizing,liu2020disparate,hu2019disparate,emelianov2020fair,immorlica2019access,garg2020standardized} usually poses the admissions problem as a classification task to predict whether an applicant is ``qualified'' or ``unqualified'' to attend their university
based on covariates given in the student's application (standardized test scores, demographic information, etc.).
When framed as a machine learning problem, the task at its core is to accept students who are qualified and reject those who are not.
However, when one widens the scope of the problem, one soon realizes that universities have finite capacity for accepting students, which creates market competition and thus strategic incentives among schools and applicants.
This latter problem is studied through a dual ML-MD lens by  \citet{emelianov2020fair}, who consider admission policies under implicit bias, and show how affirmative action in the form of group-specific admission thresholds can improve  diversity and academic merit at a capacity-constrained university.

Finally, \citet{kannan2019downstream} highlight another interesting dimension in the intersection of mechanism design, machine learning and policy: downstream effects of affirmative action. 
The paper draws upon the mechanism design literature to explore how the effects of different policy schemes propagate across education and labor when sequential decisions are made by utility-maximizing agents with potentially conflicting goals (universities vs. employers). They show that fairness notions such as equal opportunity and  (strong) irrelevance of group membership can be achieved only in the extreme case where the college does not report grades to the employer. 
Thus, the problem of intersectionality occurs again: in complex decision pipelines where different fairness metrics may be required yet it may be infeasible to satisfy all simultaneously  \citep{kleinberg2017inherent}, the question of what is an acceptable trade-off between utility maximization and various notions of fairness persists.

\subsection{Labor markets and gig economy}

Discrimination has been a perennial problem in labor markets.
Decades of research has shown that hiring decisions are subject to bias against disadvantaged communities \citep{bertrand2004emily, wenneras2001nepotism, quillian2017meta}.
More recently, techniques from both machine learning and mechanism design have been brought to bear in the labor market, and in particular, the gig economy, leading to a fresh wave of concern that the persistent discrimination found in traditional labor markets will manifest itself in new and unexpected ways.
In particular, we focus on two use cases: employee selection and employee evaluation.
Both of these use cases blend techniques from machine learning and mechanism design, and, as we will argue, it is impossible to adequately deal with issues of discrimination and bias without drawing upon ideas from both fields.

Emerging data-driven techniques for employee selection have begun to employ techniques from machine learning to evaluate and sort candidates \citep{bogen2018help,raghavan2020mitigating,sajjadiani2019using}.
While some contend that quantitative tools might help to reduce discrimination \citep{cowgill2018bias}, others warn that hiring discrimination will not be solved by machine learning alone \citep{gosh2017ai}.
However, hiring cannot be treated as a purely predictive problem; it requires consideration of factors like allocation, incentives, externalities, and competition, all of which feature more prominently in the mechanism design literature.
Consider, for example, the case of salary prediction \citep{chen2018linkedin}: platforms like LinkedIn use machine learning techniques to predict a job's salary.
While this might appear to be a straightforward application of machine learning, it creates strategic incentives that may produce unintended consequences.
If a candidate applies to a new position, their potential employer may be able to infer their current salary based on these predictions, enabling the new employer to reduce the salary they offer. Similar consequences can arise from efforts to predict a candidate's likelihood to leave a job \citep{jayaratne2020predicting,sajjadiani2019using}.
Moreover, many predictions about candidates are ultimately used in contexts where there is a limited hiring capacity.
As a result, predictions about candidates are often later used to rank or filter candidates---a type of mechanism.
To avoid the explicit consideration of demographic characteristics, efforts to ensure that candidates are treated fairly (usually through constraints similar to demographic parity) often come at the prediction stage \citep{raghavan2020mitigating}, but fail to make guarantees about the eventual outcomes produced by downstream mechanisms.
A more complete effort to prevent discrimination in algorithmic hiring pipelines must leverage the flexibility provided by machine learning to implement anti-discrimination solutions while taking into account the effects of downstream hiring mechanisms.

Beyond issues of discrimination in hiring, recent technological developments have fundamentally changed how labor markets work, particularly with regards to the gig economy, and thus led to a plethora of recent works in this space.
For example, \citet{rosenblat2017discriminating} and \citet{monachou2019discrimination} describe how mechanisms that use customer ratings to evaluate workers can internalize customers' discriminatory tastes.
 \citet{barzilay2016platform} call attention to the ways in which platform design can be used to create or reduce wage disparities. 
Similarly, \citet{hannak2017bias} document the existence of linguistic and other biases in employers' reviews for gig workers on two online labor platforms, TaskRabbit and Fiverr, and the negative effects of gender and racial bias on the number of reviews, rating, search and ranking.
 \citet{edelman2017racial} find evidence of discrimination against African-American  guests on Airbnb, highlighting the role that Airbnb's design choices play in facilitating this discrimination. 
Spurred in part by this work, Airbnb recently launched an initiative to study racial discrimination on their platform \citep{basu2020measuring}.
Crucially, this body of work combines insights from economics, mechanism design, and machine learning to better understand how discrimination can manifest in the gig economy.

\subsection{Criminal justice}

Recent popularity of the use of machine learning techniques in prescriptive settings has motivated several attempts to analyze the fairness aspects of predictive and statistical models, especially in the context of a critical application domain like criminal justice.
Unsurprisingly, relying on such models in practice can end up reinforcing underlying racial biases, as it has been shown in studies about neighbourhood surveillance \citep{ainow2019report} and recidivism prediction \citep{angwin2016machine}. 
The latter ProPublica study has raised a heated discussion 
leading many to advocate that the deployed system, independent of the larger criminal justice system in which it is situated, is plainly unfair.
While that was apparent in this instance, a rigorous explanation was not trivial; 
several responses argued that their claims of discrimination were mainly caused by differences in methodology, like the statistical measure of discrimination \citep{dieterich2016compas, flores2016false}.

Since the deployment of predictive models in the criminal justice system is a contested idea \citep{harcourt2008against}, knowledge about their potential advantages and pitfalls regarding fairness is crucial in order to perform a fruitful debate on 
their applicability.  
As already mentioned in Section 2.1, the proposed theoretical notions of fairness seem to present significant trade-offs \citep{jung2020fair, corbett2018measure} while some of them are impossible to simultaneously satisfy \citep{chouldechova2017fair,kleinberg2017inherent}.
Those contradictions naturally raise a major question regarding the criminal justice system and the automated decision-making systems within it: \textit{What do people consider truly fair?}
Since there doesn't seem to be a one-size-fits-all answer to this question, a natural step forward is a more participatory approach to the definition of (context-dependent) notions of fairness.
As discussed in Section~\ref{sec:humanperceptions}, some first approaches have been made \citep{woodruff2018qualitative, lee2018understanding, grgic2018human,  green2019disparate, srivastava2019mathematical, saha2020measuring} towards studying human perceptions of fairness but questions regarding who are the relevant stakeholders in criminal justice, what notions are more appropriate in that field and how to aggregate preferences still need to be answered.
But even under a ``perfect'' fairness definition, humans involved in the judicial decision-making process might be  inherently biased.
In this context, machine learning can be leveraged to mitigate these human biases \citep{valera2018enhancing} and mechanism design can be proven useful in studying the welfare implications and effects on inequality of decisions in criminal justice.

 Moreover, merely focusing on the task of fair recidivism prediction might be considered an oversimplification because the assessment of a ML system regarding innocence and guilt ignores both human incentives in the criminal justice pipeline and the humanity of the criminal justice system as a whole.
In the United States, a defendant only needs to prove their innocence when their case goes to trial in court.
Yet, 95\% of felony convictions in the United States are obtained through guilty pleas \citep{guiltypleaproblem}, and 18\% of known exonerees pleaded guilty or did not contest to crimes they did not commit.
Machine learning  techniques 
could be applied in conjunction with the critical perspective of mechanism design to better comprehend the racial disparities in both sentence and charge bargaining, as documented by \citet{berdejo2018criminalizing}.
It is worth noting that any theoretical techniques used to examine the criminal justice system should be wary of the common mechanism design assumption that people are rational expected utility maximizing agents, while desperation, selflessness, or fear often counter this assumption in the real world.

Though enlightening, theoretical understanding of fairness in risk assessment and its aforementioned aspects is not sufficient to suggest adopting the use of such systems.
The ultimate decision should be made by the respective stakeholders, considering the practical issues that need to be addressed \citep{koepke2018danger} and the particular context in which risk assessment tools are utilized \citep{stevenson2018assessing}.

\subsection{Health insurance markets}

Interactions between machine learning and mechanism design are salient for fairness in healthcare. For example, prior work studied how machine learning formulations may underpredict black patients' health care needs \citep{obermeyer2019dissecting}. Here, we draw attention to problems  at the intersection of the two fields in health insurance markets.

The Patient Protection and Affordable Care Act (ACA) \citep{ACA} was designed in part to defuse health-insurer incentives to refuse or avoid coverage to individuals with higher healthcare costs (i.e., selection incentives) \citep{cunningham2012risk}.
One way the ACA addresses this is through risk-adjustment based transfer payments: premiums are transferred from plans with lower expected costs to plans with higher expected costs, compensating insurers and fairly spreading costs.
Thus issues of fairness in mechanism design inhere at the policy-design level.

A key component of risk adjustment is estimating individual actuarial risk: inaccurate estimates can create selection incentives for insurers.
The Centers for Medicare \& Medicaid Services' Hierarchical Condition Category (HCC) model is a widely-used risk adjustment model \citep{cms2018risk,zink2020fair}.
The HCC predicts an enrollee's expected costs using demographic and diagnosis information; the HCC is therefore group-aware for some protected classes (e.g., sex, age), but is otherwise group-unaware, particularly to groups of enrollees with specific healthcare patterns related (but not limited) to diagnoses, treatments, and prescription-drug use.
Although there is evidence that the HCC accurately predicts expected costs for many groups of enrollees, there is also evidence that the HCC makes systematic errors for some groups  and that insurers often engage in benefit design to exploit the resulting selection incentives \citep{jacobs2015using, geruso2019screening}.

Thus, healthcare-policy designers, approaching the problem from a mechanism design perspective, encounter the lesson from fair machine learning that it is in general necessary to be group-aware.
ML practitioners also encounter the lesson from mechanism design that it is necessary to take into account the strategic behavior of stakeholders.
A natural machine-learning oriented response to the systematic error observed in the HCC could be to incorporate more information about enrollees into the model, but because the HCC data are provided by the health insurers, there are concerns that insurers or healthcare providers might then strategically upcode enrollees to favorably skew subsequent predictions \citep{cunningham2012risk}.

Recent work seeks to address these issues in risk adjustment by incorporating fairness interventions to learn a regression model that equalizes systematic error across groups. 
The proposed fair regression models can bring average predicted costs significantly more in line with average historical costs without a commensurately large penalty to the traditional evaluation metric of $R^2$ \citep{zink2020fair}.

We see each discipline's techniques applied  in a component-wise fashion towards 
a competitive health-insurance market that achieves socially optimal outcomes. 
Notably, neither discipline can independently achieve this goal: selection incentives cannot be defused without accurate risk adjustment; behavior cannot be changed by predictions without appropriately designed incentives.

\subsection{Determining creditworthiness}

The Financial Technology (FinTech) industry increasingly decides to whom a (home, business) loan should be awarded, and at what interest rate.
When one considers banks have finite liquid assets, determining an individual's creditworthiness quickly becomes one piece of a larger problem.
 \citet{saunders2019fintech} notes that FinTech can help streamline the application process for loans, among other benefits.
However, concerns including disparate impacts of disadvantaged communities, overcharging the poor, the unintelligibility of such algorithms, and protection under consumer laws emerge from the use of machine learning.
Overcharging the poor particularly appears to be in part a corollary of determining creditworthiness as a machine learning problem in isolation from mechanism design.

In determining creditworthiness, the true label (ability to pay back the loan) faces two shortcomings: first, it is only observed if the loan is given, and second, it might be a function of the given interest rate.
The Pew Research Center \citep{pew2017black} revealed 27.4\% of Black applicants and 19.2\% of Hispanic applicants were denied mortgages, compared with about 11\% of White and Asian applicants.
Moreover, when granted a loan, 39\% of Black applicants were charged an interest rate over 5\%, compared to only 28\% of White applicants.
This in turn makes repaying the loan more difficult, exacerbating financial insecurities resulting from historical financial and housing oppression, such as loan denial and redlining of neighborhoods.
Resulting from the historical imbalance of loan acceptance \citep{rambachan2019bias,liu2018delayed}, \citet{kallus2018residual} observe that algorithms might still yield ``bias in, bias out'' phenomena, even with fairness constraints, and online machine learning approaches aim to face this issue by incentivizing exploration \citep{joseph2016fair,joseph2016fairness,raghavan2018externalities,celis2019controlling,schumann2019group,kilbertus2020fair}.

As transparency increasingly becomes a legal obligation of financial institutions, such technology is particularly susceptible to disparities \citep{bartlett2019consumer}, largely because of two tensions shaped by the incentives of different stakeholders.
First, individuals who gain insight about the institutions' decision-making processes, might have disproportionate recourse abilities, based on their current financial status and access to opportunity.
Since financial institutions are typically for-profit organizations aiming to maximize their utility,  \citet{milli2019social} note that a lack of strategyproofness can disproportionately harm disadvantaged groups in the population.
The second tension arises from the fact that financial institutions are asked to find a balance between transparency towards customers and protection of their intellectual property \citep{milli2019model,barocas2020hidden}.
 \citet{tsirtsis2020decisions} argue that maximizing utility in this context may provide limited recourse to disadvantaged populations and they propose methods to counteract such disparities.
Those tensions reinforce the need to incorporate the study of incentives in automated decision-making systems before they can be effectively used in financial environments.
As  \citet{saunders2019fintech} concludes their report: \emph{The key to FinTech is: Understand first.  Proceed with caution.}

\subsection{Social networks}

Social networks have received scrutiny in the way they reinforce patterns of social inequality and discrimination \citep{mcpherson2001birds,calvo2004effects,dimaggio2011network,gundougdu2019bridging,okafor2020all}. 
Inequality at the level of individual connections is often reinforced by algorithms that use these connections for learning: in opinion diffusion \citep{fish2019gaps,ali2019discrimination,stoica2020seeding}, recommendation  \citep{stoica2018algorithmic}, clustering \citep{chierichetti2017fair,kleindessner2019fair}, and others. 
Often, such inequality arises from the individual preference for establishing new connections as well as from pre-defined communities \citep{avin2015homophily}. 
Recent papers discuss these patterns through the lens of welfare economics or equilibrium strategies, with \citet{avin2018preferential} analyzing the utility function for which preferential attachment is the unique equilibrium solution in a social network. 
Thus, understanding the incentives behind network creation patterns is crucial for designing better algorithms that learn from relational data and tackling bias at its root cause, as Section~\ref{sec:longterm} teaches us.

Beyond this, several works argue that ranking and retrieval algorithms not only reinforce existing bias, but also cause changes in people's behavior \citep{o2016weapons}. To tackle this, a recent line of work takes into consideration the post-ranking and post-recommendation effects in a game-theoretical framework, considering users as players and assigning highly ranked/recommended items to a high pay-off. The lesson from Section 3.3 of modeling individuals as rational agents has started a whole subfield in recommendation systems, starting with \citet{bahar2015economic}, who focus on finding stable equilibria for which users get the best pay-off for their desired items.  \citet{ben2018game} propose new methods, such as the Shapley mediator, to fulfill both fairness and stability conditions (as defined by mechanism design) in cases where content providers are strategic to maximize utility and assume a rational behavior of their users based on their preferences. To account for the incentives of users in post-recommendation settings, \citet{basat2017game,ben2015probability} account for users attempting to promote their own content in information retrieval, describing it as an `adversarial setting'. The main results point to an increase in general utility when accounting for such incentives, as non-strategic design presents limitations in truly fulfilling individual preferences. 
 \citet{49323} directly tackle the problem of welfare by considering recommendations as a resource to be allocated. Incorporating the preferences of the users of a social network in a fair way is thus a subsequent question. Recent works \citep{chakraborty2019equality} tackle this by adapting tools from social
choice theory, specifically, by proposing a voting mechanism called Single Transferable Vote to aggregate inferred preferences (votes) of users and achieve better recommendations. This kind of tools can be used to operate in adversarial settings, for example in non-personalised recommendation systems like Twitter or Youtube trending topics, which can be manipulated by flooding the network with bot-created content that can become viral. Methods from mechanism design can be used to protect against strategic behavior that could game the underlying machine learning system, as well as incorporate individual preferences in a meaningful way.

\section{Conclusion}\label{sec:conclusion}

While the literature is rapidly growing, many open questions  at the intersection of mechanism design and machine learning remain, motivating the need for developing a \textit{lingua franca} of fairness, identifying knowledge gaps and lessons, and ultimately bridging the two fields to work towards a fair pipeline in decision making. 

However, both communities must acknowledge  that making the pipeline ``fair'' from a technical perspective does not  mean the system is \textit{ipso facto} perfect or just.
 More interdisciplinary work  is needed beyond mechanism design and machine learning to create interventions that improve access to sociotechnical systems 
and  design  algorithms for critical application domains.

~\\\noindent\textbf{Acknowledgments.} The authors would like to thank Rediet Abebe, Itai Ashlagi, Rafael Frongillo, Nikhil Garg, Jon Kleinberg, Hannah Li, Irene Lo, Francisco Marmolejo-Cossio, Angela Zhou, and anonymous reviewers of the ACM FAccT conference for helpful comments and suggestions. This project has been part of the MD4SG working group on Bias, Discrimination, and Fairness.

This material is based upon work supported by the National Science Foundation Graduate Research Fellowship under Grants No. 1650115 (Finocchiaro), 1644869 (Maio), 1650441 (Raghavan) and 1761810 (Stoica).
Any opinions, findings, and conclusions or recommendations expressed in this
material are those of the author(s) and do not necessarily reflect the views of
the National Science Foundation. Stoica acknowledges support from the J.P. Morgan AI research fellowship. Monachou acknowledges support from the Krishnan-Shah Fellowship and the A.G. Leventis Foundation Grant. 
Patro is supported by a fellowship from Tata Consultancy Services Research.
Tsirtsis acknowledges support from the European Research Council (ERC) under the European Union’s Horizon 2020 research and innovation programme (grant agreement No. 945719).

\small
\bibliographystyle{plainnat}

\bibliography{refs}

\end{document}